\documentclass{article}

\usepackage{arxiv}

\usepackage[utf8]{inputenc} 
\usepackage[T1]{fontenc}    
\usepackage{hyperref}       
\usepackage{url}            
\usepackage{booktabs}       
\usepackage{amsfonts}       
\usepackage{nicefrac}       
\usepackage{microtype}      
\usepackage{lipsum}

\usepackage{graphicx}      
\usepackage{dcolumn}
\usepackage{bm}
\usepackage{amsmath}    
\usepackage{multirow}
\usepackage{bbm}
\usepackage{xcolor}
\usepackage{soul}
\newcommand{\mathcolorbox}[2]{\colorbox{#1}{$\displaystyle #2$}}
\usepackage{relsize}

\title{Interarea Oscillations \& Chimera in Power Systems}

\author{
  Pratik K.~Bajaria\\
  Research Scholar, Department of Electrical Engineering,\\ 
  Veermata Jijabai Technological Institute,\\
  Mumbai 400019, Maharashtra, India.\\
  \texttt{pkbajaria\_p15@ee.vjti.ac.in} \\
   \And
  Sushama R.~Wagh\\
  Faculty, Department of Electrical Engineering,\\
  Veermata Jijabai Technological Institute,\\
  Mumbai 400019, Maharashtra, India.\\
  \texttt{srwagh@ee.vjti.ac.in} \\
  \And
  Navdeep M.~Singh\\
  Faculty, Department of Electrical Engineering,\\
  Veermata Jijabai Technological Institute,\\
  Mumbai 400019, Maharashtra, India.\\
  \texttt{nmsingh@ee.vjti.ac.in} \\
}

\begin{document}
\maketitle

\begin{abstract}
This paper proposes a novel second order mathematical model in the Kuramoto framework to simulate and study low frequency oscillations in power systems. This model facilitates better understanding of the complex dynamics of a power network. A standard four generator power system with all-to-all connectivity is considered and results obtained from the proposed model are verified. It is shown, that the model simulates various properties related to low frequency oscillations in power systems which presently are obtained through small-signal analysis. Further, we provide analogy to blackouts in a power grid, by emulating chimera behavior and thereby discuss bifurcation analysis of the proposed model.
\end{abstract}


\section{INTRODUCTION}

\par Electrical power systems are complex engineering networks which are crucial to the present day infrastructure. A power network comprises of blocks consisting of numerous interconnected sub-systems, making them challenging to analyze and understand. With the continuous increase in electricity demand and the trend for more interconnections, an issue of concern is the mitigation and analysis of low-frequency interarea oscillations. Oscillations associated with individual generators in a power plant are called local mode oscillations typically ranging from 0.7-2.0Hz \cite{rogers2012power, klein1991fundamental}. The stability of these oscillations characterized as intraarea (same area) and interarea (across areas) \cite{rogers2012power, klein1991fundamental}. These oscillations between the generators which are inherent to power systems require appropriate mathematical models and techniques for their analysis.
\par Kuramoto-type models have been widely used to study the dynamics of a power system network through swing equations \cite{filatrella2008analysis}. It must be noted though that a power system network has an added second order term due to generator inertia and are only similar to Kuramoto model. Power dissipation terms that arise in the swing equation model are absent in conventional Kuramoto model, which can be shown existent by few mathematical adjustments. In power systems, globally coupled phase oscillators of Kuramoto form have been viewed as electromechanical generators mutually coupled to deliver load power. A conventional second order Kuramoto-type oscillator can be written as follows,
\begin{equation}
J_i\ddot{\delta}_i+d_i\dot{\delta}_i=\omega_i+\sum_{j\neq i,j=1}^{n}k_{ij}sin(\delta_j-\delta_i), \ \ \ i \in \{1,\ldots,n\},
\label{kuramoto}
\end{equation}
where $\delta_i$ is angular position of rotor with respect to the synchronously rotating reference frame (for consistency, all the angles throughout the paper are in radians), $J_i$ inertia in kgm$^2$, $d_i$ damping and $\omega_i$ is a natural frequency chosen from an appropriate distribution $g(\omega)$ of $i$-th oscillator. $[k_{ij}]; i,j\in \{1,\ldots,n\}$ is the matrix of coupling constants and $n$ defines the number of oscillators. The standard Kuramoto-type equation assumes value of coupling constants $[k_{ij}]$ to be always positive and symmetric. In this paper, author explores the mapping between a power grid and Kuramoto oscillators.
\par Interarea oscillations emerge when two areas having independent sets of power generators experience supply-demand imbalance. The generators in individual areas are observed to beat against each other with frequencies ranging from 0.1-0.8Hz, classified as low frequency interarea oscillations in a power grid. These oscillations can be visualized as two large generators trying to desynchronize each other in the event of supply-demand balance being achieved in each individual area. The above phenomena is analyzed using small-signal or modal analysis \cite{klein1991fundamental}, however additionally it would be  advantageous to have a nonlinear (large-signal) model to capture the different behaviors and effects of these oscillations. We propose a novel `conformist-contrarian' (inspired from first order framework discussed in \cite{hong2011kuramoto}) second order Kuramoto-type model (henceforth, referred to as CC-Kuramoto) which captures the in-phase (intraarea) and the anti-phase (interarea) oscillations in a power system.
\par The motivation behind developing such a model is to address some of the challenges related to modeling low frequency oscillations in power systems \cite{kundur2002small}. Conventionally, small-signal analysis and damping control is used by power system engineers to assure system stability at planning stage and thereby execution. It has to be noted though, that over the years software packages on these design/analysis have become computationally efficient in terms of execution time, but still bears significant computation cost for near real-time implementation. Some key challenges related to models for power systems, identified from the literature are as follows:
\begin{enumerate}
    \item The major problems related to power system oscillations are of perturbed damping of overall system which are regulated conventionally using power system stabilizers. These oscillations are identified using eigen value analysis which are computationally costly.
    \item In cases when power transfer needs to be increased or decreased, the groups of individual generators in the source and sink sides are dispatched in order of their sensitivities to the critical modes with respect to the output of these generators. These in turn assist to increased power levels without adding any further damping control actions. Computation of these critical modes and calculation of generator sensitivities to it in real-time is difficult.
    \item Modeling such behaviors is not an easy task due to observed in-coherency from planning to actual implementations in the past. Details of major equipment and inclusion of newer loads like those of induction motors is still not simple in small-signal models.
    \item On the similar lines, a $-$ sync behavior in system characteristics has tendency to mitigate homogeneous power oscillations in an interarea setup \cite{shim2017synchronization}. The behavior and analysis of interarea oscillations in a nonlinear form by adding periodic disturbances to the major parameters that have significant impact \cite{mao2008nonlinear}. It would be advantageous, if a model can integrate these results through simple modifications.
\end{enumerate}
\par Apart from fabricating a perfect model that can overcome above mentioned complexities, a power system engineer looks for a model that can provide significant inferences. With increasing vulnerability of modern power systems due to inclusion of various ancillary services it is important to study the settings that might lead to partial stability or instability. It must be noted though, that power grids are not simple physical network of transmission lines and are deeply impacted by its structural as well as dynamical interactions. Thus, a dynamic redesign/modification of existent power network is not possible, as it can be a major limiting factor in optimizing synchronization. With these constraints as reference, we show occurrence of various stable, partially stable and unstable states via tuning of system parameters, and avoid fiddling with the structure. It is observed that these parameters beyond a certain threshold lead to randomization of steady-state equilibrium points thereby existence of a chaotic behaviour. The same power grid setup (and some other complex systems in nature) shows a state of partial stability by clustering themselves into islands of synchronised and de-synchronised oscillators, commonly referred to as chimera in literature \cite{abrams2004chimera}. Hence, we emulate the existence of these chimera state behaviors and correlate them with blackouts with islanding commonly seen in power grids \cite{nerc9209}.
\par Nonlinear modes associated with instabilities have been analysed and discussed in the literature \cite{susuki2011nonlinear, susuki2009global, parrilo1999model} related to power grid synchronization. These provide an informative decomposition of nonlinear oscillations when the network looses synchrony. Sufficient conditions for synchronization, obtained via perturbation analysis for non-uniform Kuramoto oscillators are also widely studied \cite{dorfler2012synchronization, dorfler2011critical}. It must be noted though, that the conditions attained in previous studies maintain homogeneity in system parameters, whereas we obtain conditions on various power grid parameters and hence heterogeneity. On the similar lines, effects of heterogeneity on power grid networks discussed in \cite{motter2013spontaneous} inspire formulation of Kuramoto-type framework and extend it to practical blackout scenarios. Thus, in this work we take a standard example from power systems to show that chimera behaviors can lead to blackouts and can be correlated to a distributed grid. To summarize, major contributions of this work are as follows. We propose a nonlinear model for analysis of power systems in simplistic form to avoid discussed computational complexities. A practical example from power systems is used to correlate and showcase advantages of the proposed model. It has been shown, that the model not only provides information about the nominal states but also existence of chimera behaviors in power systems. This could help site engineers to take actions apriori or equip with necessary tools at the right time.
\par The manuscript is divided into two parts. \emph{Part-\uppercase\expandafter{\romannumeral 1}}: We first start with modeling a standard power grid in Kuramoto form. Then, gradually move towards addressing complexities discussed before and how it can be easily incorporated in a large scale (nonlinear model) using analogy to a standard physics example. Next, a standard power systems network to study interarea oscillations is considered and results are verified using computer simulations. \emph{Part-\uppercase\expandafter{\romannumeral 2}}: A detailed bifurcation analysis is performed on the proposed model parameters in order to analyse the system stability. Finally, we emulate chimera behavior \cite{abrams2004chimera} commonly referred to in the literature and discuss its implications in a power network.


\section{Part - \uppercase\expandafter{\romannumeral 1}: Power Network and Kuramoto Oscillators}\label{mod1}

\subsection{Mathematical model of Power Grid}\label{kura}

\par The basic elements of a power grid consists of active generators and passive machines/loads. The generator converts some source of energy into electrical power which is produced by the prime mover of the generator with the frequency close to the standard or natural frequency $ \Omega $ of an electrical system. All generators in a power grid can be looked upon as set of synchronous machines rotating at synchronous frequency $\Omega$, with the stator windings of the generator delivering electrical power to the grid. Any power generator in a power system is described by a power balance equation of the form,
\begin{equation}
P_{accumulated}+P_{dissipated}=P_{source}-P_{transmitted},
\label{sumpower}
\end{equation}
where $P_{source}$ is the rate at which the energy is fed into the generator at frequency $\Omega$ (i.e., $2\pi\Omega=50$Hz). Hence, the phase angle $\theta_i$ at the output of the $i$-th generator in stationary frame is then given by,
\begin{equation}
\theta_i=\Omega t +\delta_i.
\label{theta_eq}
\end{equation}
$P_{accumulated}$ is the rate at which kinetic energy is accumulated by the generator:
\begin{equation}
P_{accumulated} = \frac{1}{2} J_i \frac{d}{dt}(\dot{\theta}_i)^2,
\label{pacc}
\end{equation}
where $ J_i $ is the moment of inertia of the $i$-th generator in kgm$^{2}$. For the sake of simplicity, we assume identical machines (i.e., $J_i=J$). $P_{transmitted}$ is the power transmitted from generator $i$ to $j$ with phase difference, $\Delta \theta_{ij}=\theta_j-\theta_i \neq 0$. 
\begin{equation}
P_{transmitted}=-P_{max}sin(\Delta\theta_{ij}).
\label{ptrans}
\end{equation}
$P_{max}$ being maximum electrical power input in watts. The dissipated power ($P_{dissipated}$) with $K_D$ the dissipation constant of the prime mover in Ws$^2$/rad$^2$, can be expressed as: 
\begin{equation}
P_{dissipated}=K_D (\dot{\theta}_i)^2.
\label{pdiss}
\end{equation}
Since, all the generators share common frequencies $\Omega$, $\Delta \theta_{ij} = \Delta \delta_{ij} = \Phi_{ij}$. Substituting \eqref{pacc}, \eqref{ptrans} and \eqref{pdiss} in \eqref{sumpower}, following can be computed,
\begin{equation}
P_{source} = J \ddot{\theta}_i\dot{\theta}_i + K_D(\dot{\theta}_i)^2 - P_{max}sin(\Phi_{ij}).
\label{pgen}
\end{equation}
Differentiating \eqref{theta_eq} with respect to time and further double differentiating it; thereby assuming perturbations around the synchronous frequency being very small, $i.e., \dot{\delta}_i \ll \Omega$, \eqref{pgen} can be approximated as, 
\begin{equation}
P_{source} \cong J\Omega\ddot{\delta}_i+[J\ddot{\delta}_i+2K_D\Omega]\dot{\delta}_i+K_D\Omega^2-P_{max}sin(\Phi_{ij}).
\label{pgen2}
\end{equation}
Under practically relevant assumptions, the coefficient of first derivative is constant and neglecting acceleration terms, as well as knowing that the rate at which the energy is stored in kinetic term is much lower as compared to rate at which energy is dissipated in friction, \eqref{pgen2} is reduced to,
\begin{equation}
J\Omega\ddot{\delta}_i=P_{source}-K_D\Omega^2-2K_D\Omega\dot{\delta}_i+P_{max}sin(\Phi_{ij}).
\label{pgen3}
\end{equation}
Now, using the fact that, $P_{max}=E_iE_j\left|Y_{ij}\right|$; $E_i$ being internal voltage of $i$-th generator, $Y_{ij}$ the Kron reduced admittance matrix denoting maximum power transferred between generators \cite{dorfler2012synchronization} and choosing $P_{m,i}=P_{source}-K_D\Omega^2$, where $P_{m,i}$ is the mechanical power input,
\begin{equation}
J\Omega\ddot{\delta}_i=P_{m,i}-E^{2}_{i} \Re(Y_{ii})-2K_D\Omega\dot{\delta}_i
+\sum_{j\neq i,j=1}^{n}E_iE_j \left| Y_{ij}\right|sin(\Phi_{ij}).
\label{pgen5}
\end{equation}
Dividing both sides by $J\Omega$,
\begin{equation}
\ddot{\delta}_i=\left[\frac{P_{m,i}}{J\Omega}-\frac{E^{2}_{i} \Re(Y_{ii})}{J\Omega}\right]-\frac{2K_D}{J}\dot{\delta}_i
+\sum_{j\neq i,j=1}^{n}\frac{E_iE_j \left| Y_{ij}\right|}{J\Omega}sin(\Phi_{ij}).
\label{pgen6}
\end{equation}
\eqref{pgen6} can be rewritten as follows,
\begin{equation}
\ddot{\delta}_i=\omega_i-\alpha \dot{\delta}_i+\sum_{j\neq i,j=1}^{n}k_{ij}sin(\delta_j - \delta_i),
\label{pkura}
\end{equation}
where $\alpha=\frac{2K_D}{J}$ is the dissipation constant, coupling constant $[k_{ij}] = \frac{E_iE_j \left| Y_{ij}\right|}{J\Omega}$ and natural frequency $\omega_i=\left[\frac{P_{m,i}}{J\Omega}-\frac{E^{2}_{i} \Re(Y_{ii})}{J\Omega}\right]$. From a graph theoretic viewpoint $[k_{ij}]$ can be seen as a weighted laplacian matrix with $k_{ij}=0$ when generators are not connected to each other and $ k_{ij} \geq \frac{(\omega_{max}-\omega_{min})n}{(2(n-1))} $ otherwise (i.e., assumed to be greater than critical coupling, to ensure steady state synchronization \cite{dorfler2012synchronization}). It can be observed that, \eqref{pkura} has the same form as a second order Kuramoto oscillator model \cite{dorfler2012synchronization}.

\subsection{CC-Kuramoto Model for Interarea Oscillations}\label{model}
\par
\begin{figure}[t!]
\centering
\begin{tabular}{cc}
\includegraphics[height=4.0cm,width=3.9cm]{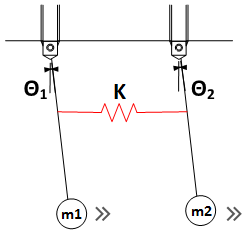}&
\includegraphics[height=4.0cm,width=4.5cm]{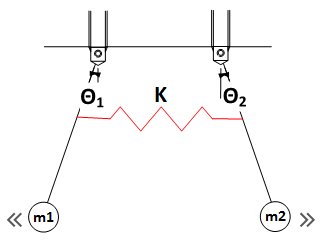}\\
(a) $\theta_1=\theta_2$ &(b) $\theta_1=-\theta_2$
\end{tabular}
\caption{Pendulums coupled by a spring, oscillating in two equilibrium modes. (a) Oscillations in in-phase mode. (b) Oscillations in anti-phase mode.}
\label{fig:pend}
\end{figure}

 Further, we extend \eqref{pkura} to spring coupled oscillators. In Kuramoto oscillators or spring coupled pendulums for that sake; the coupling term introduces restoring forces on the oscillators. Considering the special case, when there is no energy transfer between oscillators, either of in-phase or anti-phase  steady state oscillations may exist (as observed in a spring coupled pendulum - Figure \ref{fig:pend}). For the in-phase oscillations, the restoring forces are zero implying absence of the coupling term, which is not a tangible explanation for a coupled system in practice because there would always be some sort of restoring forces present in a coupled system. On the other hand, in the case of anti-phase oscillations the spring keeps contributing restoring forces whilst the energy transfer is zero \cite{feynman2011feynman}. This also explains Huygens observations \cite{bennett2002huygens} and is a valid template for modeling interarea oscillations in power systems.
\par Any oscillator of the form given in \eqref{pkura}, assuming damping/dissipation constant to be zero, with identical natural frequencies ($\omega_i=\omega $) and $H(\Phi_{ij})=\sum_{j\neq i,j=1}^{n}k_{ij}sin(\Phi_{ij})$ and $\Phi_{ij}=(\delta_j-\delta_i)=(\theta_j-\theta_i)$, can be written as,
\begin{equation}
\ddot{\delta}_i=\omega+H(\Phi_{ij}),
\label{axonal}
\end{equation}
and thereby, following can be deduced,
\begin{equation}
\ddot{\Phi}_{ij}=H(-\Phi_{ij})-H(\Phi_{ij})=-2H(\Phi_{ij}).
\label{axonal2}
\end{equation}
The above \eqref{axonal2} has fixed points $\Phi_{ij}=\beta(2\pi)$ or $\Phi_{ij}=(2\beta-1)\pi; \forall \beta \in \mathbb{Z}$ which are respectively the in-phase and anti-phase modes of the oscillator. Linearizing \eqref{axonal2} about its fixed points,
\begin{equation}
\begin{split}
\ddot{\Phi}_{ij}&\approx \left[ -2\frac{\partial H(\Phi_{ij})}{\partial \Phi_{ij}}\Bigr|_{\substack{H(\Phi_{ij})=0}} \right] \Phi_{ij},\\
&\approx \left [-2k_{ij}cos(\Phi_{ij})\Bigr|_{\substack{H(\Phi_{ij})=0}} \right ]\Phi_{ij}.
\end{split}
\label{axonal3}
\end{equation}
Thus, from \eqref{axonal3} it can be deduced that in-phase solution $\Phi_{ij}=0$ is synchronizing and stable, whereas anti-phase solution $\Phi_{ij}=\pi$ is desynchronizing and unstable. These results are similar to small-signal stability analysis performed by linearizing the nonlinear power system dynamics \cite{klein1991fundamental}. Hence, next we integrate these equilibrium/critical modes directly in the nonlinear dynamics of Kuramoto model \eqref{pkura}. 
The in-phase or `conformist' model of Kuramoto oscillators can be given as follows,
\begin{equation}
    \ddot{\delta}_i=\omega_i-\alpha_i \dot{\delta}_i \ \mathcolorbox{yellow}{\mathlarger{\mathlarger{+}}}\sum_{j\neq i,j=1}^{n}k_{ij}sin(\delta_j - \delta_i),
    \label{conf}
\end{equation}
whereas, an anti-phase or `contrarian' Kuramoto model can be obtained by replacing $H(\Phi_{ij})$ with $-H(\Phi_{ij})$ in \eqref{axonal} to give,
    \begin{equation}
    \ddot{\delta}_i=\omega_i-\alpha_i \dot{\delta}_i\ \mathcolorbox{green}{\mathlarger{\mathlarger{-}}}\sum_{j\neq i,j=1}^{n}k_{ij}sin(\delta_j - \delta_i).
    \label{cont}
\end{equation}
\par Linearisation of the `contrarian' model \eqref{cont} on the lines of \eqref{axonal3} yields $\ddot{\Phi}_{ij}\leq 0$ for anti-phase modes and otherwise for in-phase modes. Thus, in the `contrarian' model the anti-phase mode is stable and in-phase mode is unstable. All oscillations in physical systems in general and power systems in particular are weighted sum of in-phase and anti-phase modes. Hence, to study the oscillations in power systems, we propose the CC-Kuramoto model of coupled oscillators given as,
        \begin{equation}
        \begin{split}
        \ddot{\delta}^{a_1}_i=\omega_i-\alpha_i \dot{\delta}^{a_1}_i&\ \mathcolorbox{yellow}{\mathlarger{\mathlarger{+}}}\sum_{j\neq i,j=1}^{p}k_{ij}sin(\delta_j^{a_1} - \delta^{a_1}_i)\\
        &\ \mathcolorbox{green}{\mathlarger{\mathlarger{-}}}\sum_{j=p+1}^{n}k_{ij}sin(\delta^{a_2}_j - \delta^{a_1}_i),\\
        \ddot{\delta}^{a_2}_i=\omega_i-\alpha_i \dot{\delta}^{a_2}_i&\ \mathcolorbox{yellow}{\mathlarger{\mathlarger{+}}}\sum_{j\neq i,j=p+1}^{n}k_{ij}sin(\delta_j^{a_2} - \delta^{a_2}_i)\\
        &\ \mathcolorbox{green}{\mathlarger{\mathlarger{-}}}\sum_{j=1}^{p}k_{ij}sin(\delta^{a_1}_j - \delta^{a_2}_i),\\
        \end{split}
        \label{interarea}
        \end{equation}
        where without loss of generality we assume $\delta_i^{a_c} \in \big[ \delta_1^{a_1},\delta_2^{a_1},\delta_3^{a_2},\delta_4^{a_2}\big]$, $ p $ set of generators in area 1 ($a_1$) and $ (n-p) $ generators in area 2 ($a_2$).
        
\par Further, a CC-Kuramoto model settles into one of the three type of states, depending upon the system parameters and initial conditions: Incoherent state - a state of complete desynchronization, $\pi$-state - when two groups of coherent oscillators are separated by phase difference of $\pi$ radians and the Travelling wave state - where the two coherent groups are apart by a phase difference less than $ \pi $ radians. Not only do the above states exhibit rich dynamical behavior but other interesting outcomes arise in the process of transition between these states. The direct relation between CC-Kuramoto and the power system network facilitates the study of complex dynamics arising in power networks.

\subsection{Case study: Classical two-area four-machine system}\label{study}

\par
\begin{figure}[t!]
\centering
\includegraphics[height=3.3cm,width=8cm]{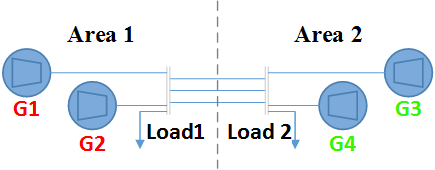}
\caption{Single line diagram of a two-area four-machine power system.}
\label{fig:sld}
\end{figure}

 In order to validate the proposed model, we use a classical two-area four-machine power system commonly referred to for interarea oscillation analysis \cite{klein1991fundamental}. The system is symmetric; consisting of two identical areas connected through a relatively weak tie ($J_i=J=0.4$kgm$^2$, $\alpha_i=\alpha=0.125$). Each area includes two synchronous generators with equal power output. The single line diagram of the system considered is as shown in Figure \ref{fig:sld}, 
\par All loads are represented as constant impedances. The tie-line impedance was varied by changing the number of tie circuits in service. Power transfer between two areas is emulated, either by an uneven distribution of generation between the areas, or by an uneven split of the total system loads. The combinations for tie-line power flow are given in Table \ref{data}.
\begin{table}[b!]
\centering
\caption{Load and Tie-Line Power of Test System}
\label{data}
\begin{tabular}{|l|l|l|l|l}
\cline{1-4}
     & \multicolumn{2}{c|}{Generation/Load (MW)} & \multicolumn{1}{c|}{\multirow{2}{*}{\begin{tabular}[c]{@{}c@{}}Power flow from Area 1 \\ to Area 2 (MW)\end{tabular}}} &  \\ \cline{1-3}
       & \multicolumn{1}{c|}{Area 1}            & \multicolumn{1}{c|}{Area 2}           & \multicolumn{1}{c|}{}                                                                                                 &  \\ \cline{1-4}
Case 1 & 1400/1367            & 1400/1367              & \multicolumn{1}{c|}{0}                                                                                                                   &  \\ \cline{1-4}
Case 2 & 1400/967              & 1450/1767            & \multicolumn{1}{c|}{400}                                                                                                                 &  \\ \cline{1-4}
\end{tabular}
\end{table}

\par The number of tie-line in service is two and the transfer level along the tie-line of the two areas varies from 0 to 400 MW due to the variation of load levels in the two areas. Case 1 relates to no power transfer between two areas. On the other hand, the event of power transfer between areas has been designated as Case 2.
\begin{align}
K_{Case 1}=\begin{bmatrix}0 & 1.9689 & 0.1766 & 0.1782\\ 
                             1.9689 & 0 & 0.1782 & 0.1801\\
                             0.1766 & 0.1782 & 0 & 1.9363\\
                             0.1782 & 0.1801 & 1.9363 & 0 \end{bmatrix}
\label{Kmatrix1}
\end{align}
\begin{align}
K_{Case 2}=\begin{bmatrix}0 & 2.5960 & 0.2130 & 0.2151\\ 
                             2.5960 & 0 & 0.2151 & 0.2171\\
                             0.2130 & 0.2151 & 0 & 1.7214\\
                             0.2151 & 0.2171 & 1.7214 & 0 \end{bmatrix}
\label{Kmatrix2}
\end{align}
The coupling matrix $[k_{ij}]$ is represented as (\ref{Kmatrix1}) for Case 1 and (\ref{Kmatrix2}) for Case 2. Elements of coupling matrix $[k_{ij}]$ are derived from $k_{ij}=\frac{E_iE_j\left| Y_{ij} \right|}{J\Omega} $. Natural frequencies are calculated using $\omega_i=\left[\frac{P_{m,i}}{J\Omega}-\frac{E^{2}_{i} \Re(Y_{ii})}{J\Omega}\right]$ and are shown in Table \ref{freq_data}. For simplicity, we assume $\Omega=1$Hz.
\begin{table}[b!]
\centering
\caption{Natural frequencies (in rad/s)}
\label{freq_data}
\begin{tabular}{lllll}
\hline
\multicolumn{1}{|l|}{}       & \multicolumn{1}{c|}{$\omega_1$ } & \multicolumn{1}{c|}{$\omega_2$ } & \multicolumn{1}{c|}{$\omega_3$ } & \multicolumn{1}{c|}{$\omega_4$ } \\ \hline
\multicolumn{1}{|l|}{Case 1} & \multicolumn{1}{c|}{17.5290}                         & \multicolumn{1}{c|}{17.7923}                         & \multicolumn{1}{c|}{17.5640}                         & \multicolumn{1}{c|}{17.8285}                         \\ \hline
\multicolumn{1}{|l|}{Case 2} & \multicolumn{1}{c|}{16.8882}                         & \multicolumn{1}{c|}{17.1532}                         & \multicolumn{1}{c|}{17.7931}                         & \multicolumn{1}{c|}{18.0629}                         \\ \hline                                       
\end{tabular}
\end{table}

\begin{figure*}[t!]
\begin{tabular}{ccc}
\includegraphics[height=4.0cm,width=4.8cm]{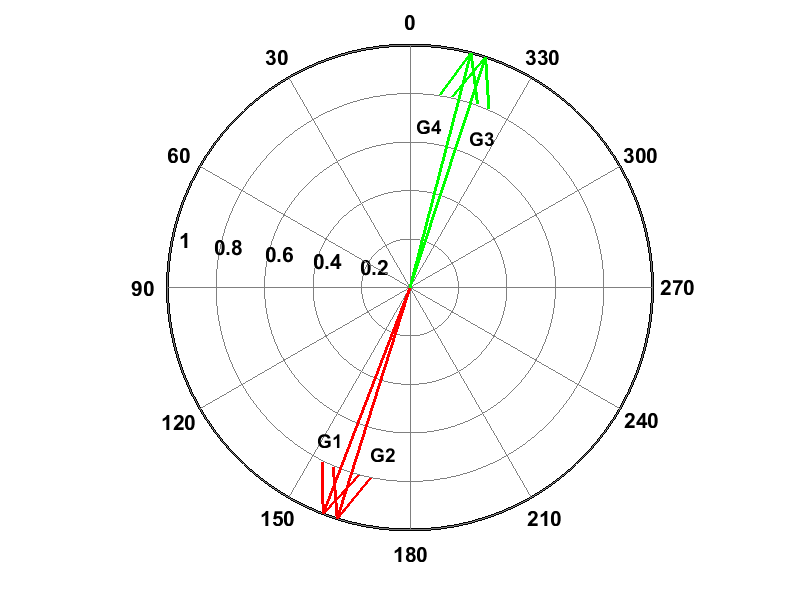}&
\includegraphics[height=4.0cm,width=4.8cm]{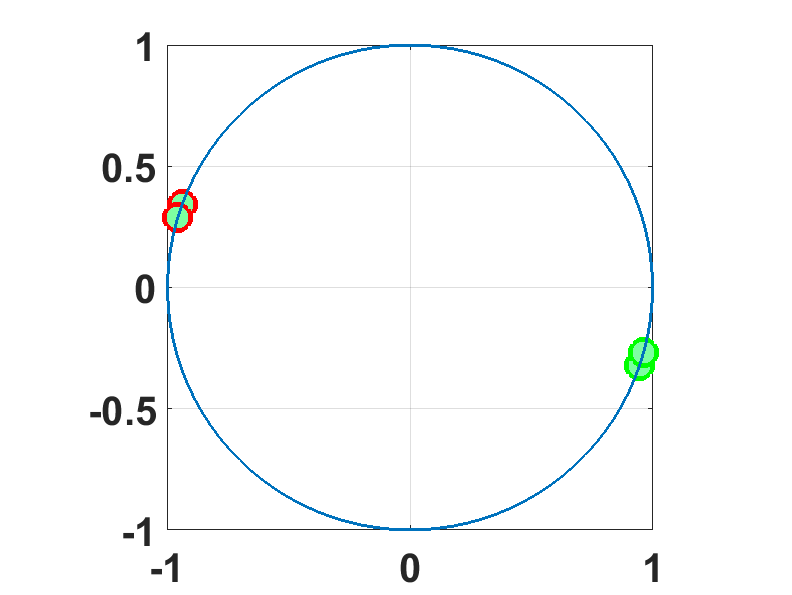}&
\includegraphics[height=4.0cm,width=6cm]{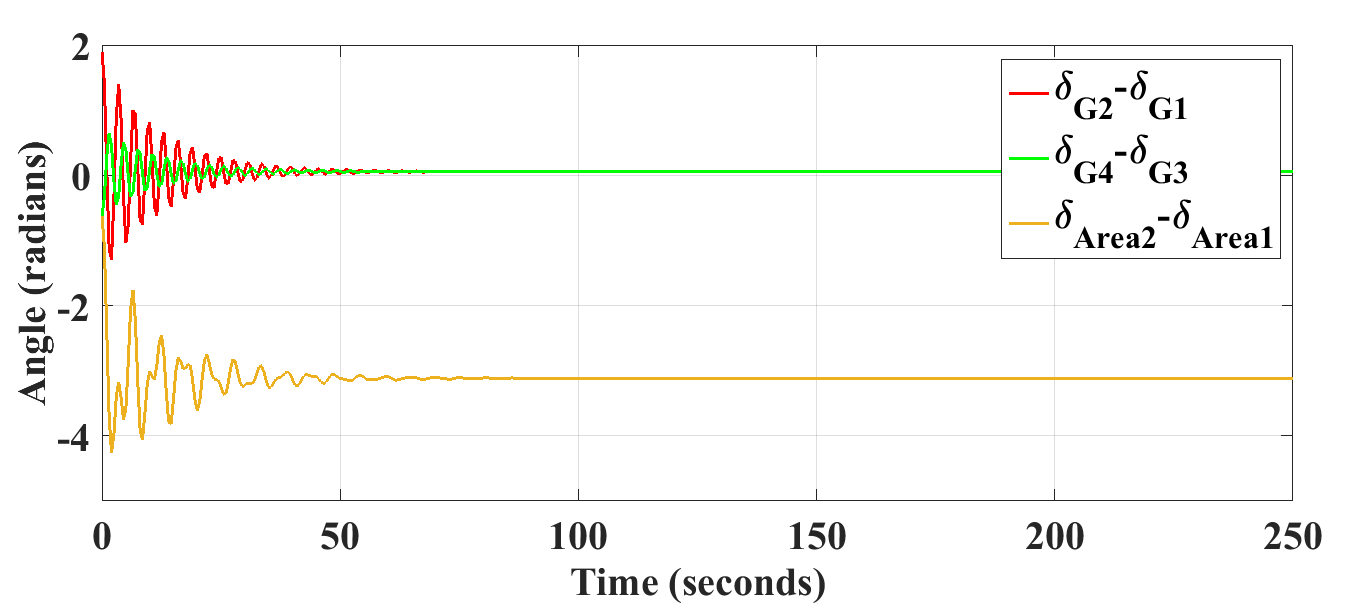}\\
(a)&(b)&(c)\\
\includegraphics[height=4.0cm,width=4.8cm]{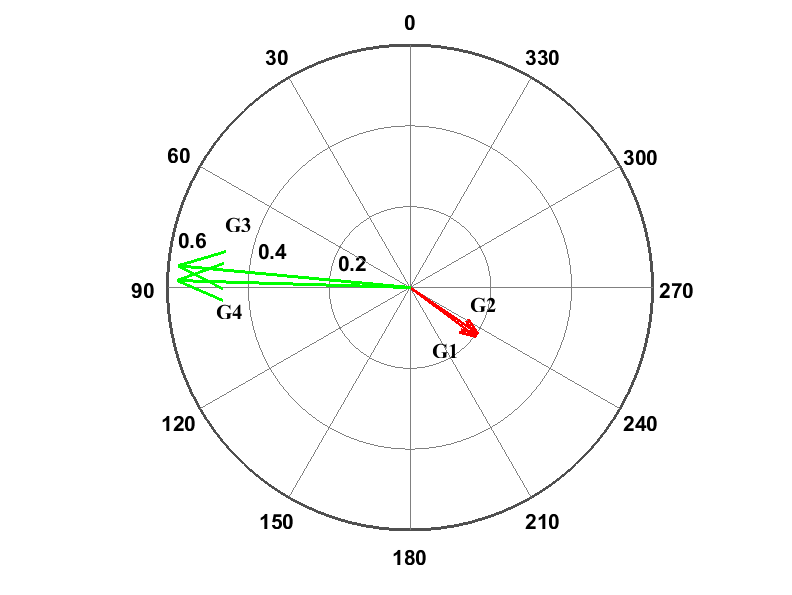}&
\includegraphics[height=4.0cm,width=4.8cm]{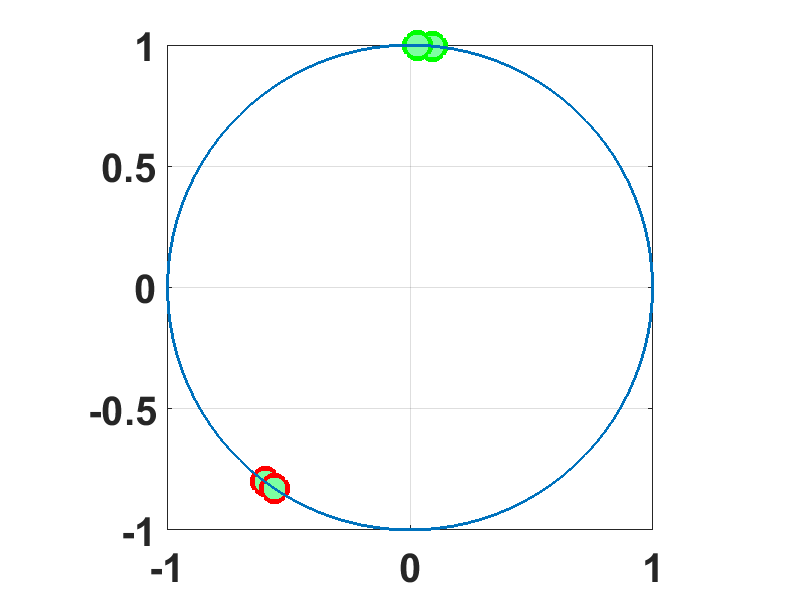}&
\includegraphics[height=4.0cm,width=6cm]{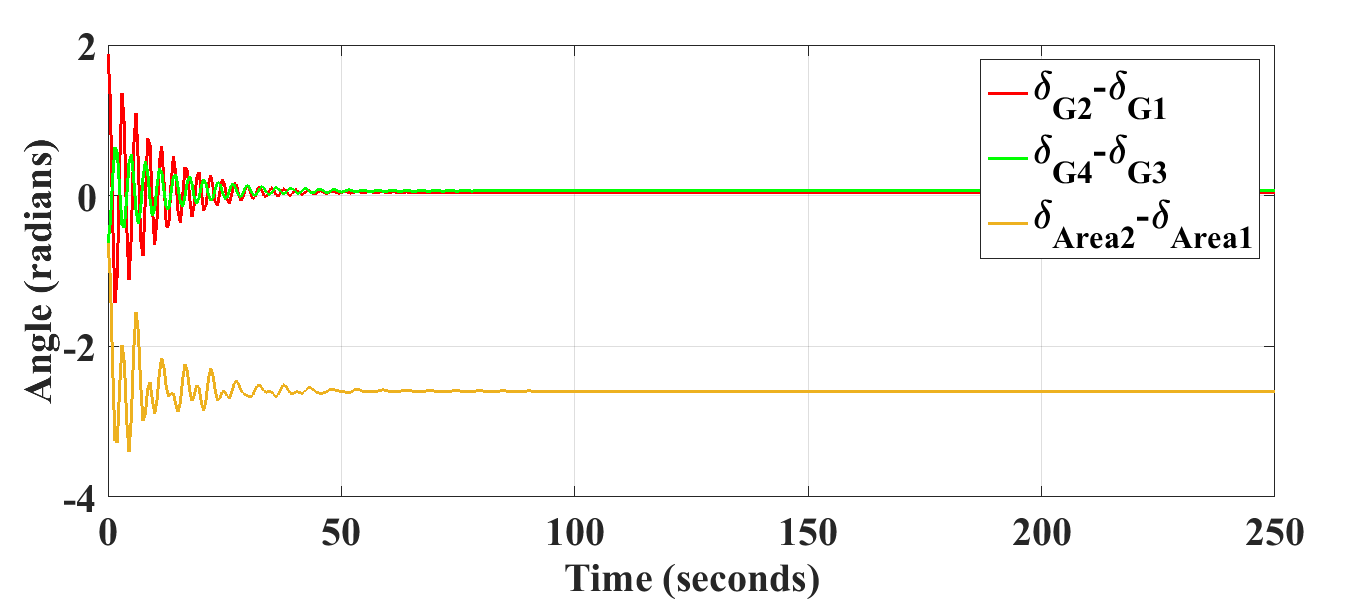}\\
(a)&(b)&(c)
\end{tabular}
\caption{Case 1 (first row from the top) - Dynamics of the proposed model for interarea oscillations, when no power is transferred. (a) Compass plot. (b) Circle plot. (c) Time series plot. Case 2 (second row from the top) - Dynamics of the proposed model for interarea oscillations, when power is transferred from area 1 to area 2. (a) Compass plot. (b) Circle plot. (c) Time series plot.}
\label{fig:case2}
\end{figure*}

\par The model proposed in \eqref{interarea} was solved using MATLAB and following observations were made.
\par \textit{Case 1}: The generators oscillate anti-phase ($-3.12 \approx -\pi$radians) in interarea and in-phase ($0.06 \approx 0$radians) intraarea. To validate the results, we provide compass plots of normalized eigen modes, circle plot as well as time-domain plots of generators as shown in Figure \ref{fig:case2}. The compass plots were obtained by using the steady-state vectors: $\vec{c}_i(t) = \big (\delta_j-\delta_i \big )_{rms} \angle \delta_i(t) \sim \big ( \delta_j-\delta_i \big )_{rms} \left( \sum^{n}_{i=1} e^{\lambda_i t} u_i v^{T}_i \delta_i(0) \right)$, where $u_i$ is the normalized left eigen-vector, $v_i$ is the normalized right eigen-vector and $\lambda_i$ are the eigen-values of the linearized system. It can be seen that the compass plot of steady state vectors show behavior similar to normalized eigen modes obtained by small-signal analysis performed traditionally. 
\par \textit{Case 2}: In the case when power is transferred between areas, the phase difference between interarea generators were observed to be $-2.6$radians (i.e., $  \neq-\pi$) and $ 0.05 $radians in intraarea, as shown in Figure \ref{fig:case2}.
The results are compared with \cite{klein1991fundamental}, providing validity to the proposed model.

\section{Part - \uppercase\expandafter{\romannumeral 2}: Partial Stability in Power Systems}

\par In this section, we study the bifurcation analysis of the proposed CC-Kuramoto model in order to understand the effect of the design parameters on the stability of a power network. In order to do so, we first analyse some of the characteristics nodes in power systems.

\subsection{Equal Area Criteria in Power Systems}

\begin{figure}[t!]
\centering
\includegraphics[height=7.2cm,width=8.2cm]{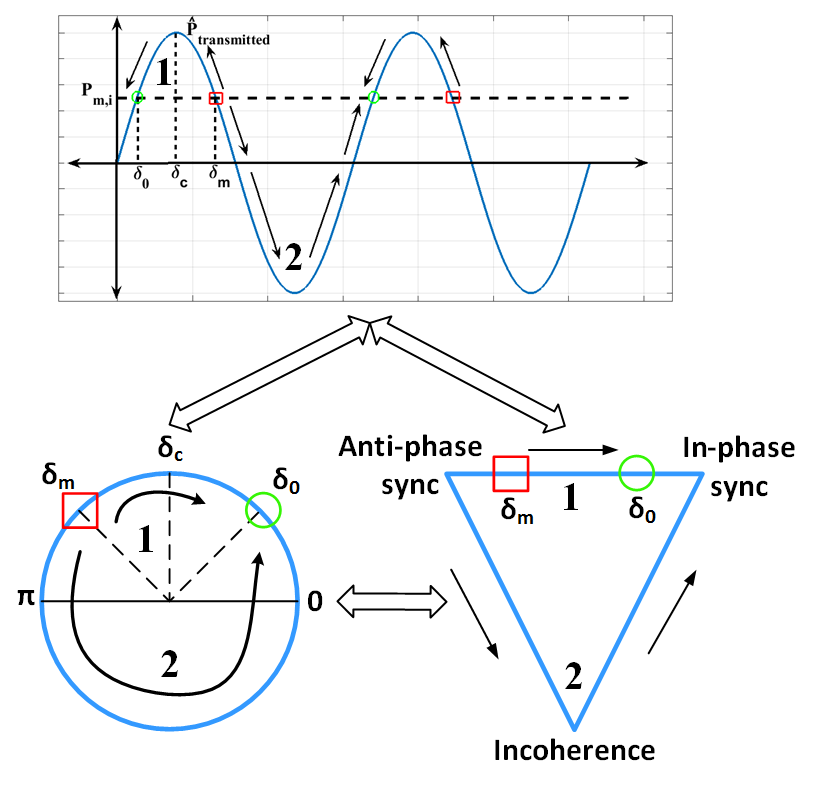}
\caption{Equivalence of equal area criterion to equilibrium spaces. Green circles denote `sink' nodes where system achieves equilibrium. Red squares denote `source' node where system attains acceleration and hence increase in accumulated power.}
\label{fig:equal_area_martens}
\end{figure}

\par In power systems, the equal area criterion is a ``graphical technique used to examine the transient stability of the machine systems (one or more than one) with an infinite bus". The areas under the curve of a power angle diagram are equated across to calculate effective acceleration/deceleration thereby comment on the stability of the system. For instance, consider \eqref{interarea}, rewriting in terms of mechanical and electrical power interactions,

\begin{equation}
    \ddot{\delta}_i=P_{m,i}-\hat{P}_{transmitted},
    \label{equal_area}
\end{equation}

where $\hat{P}_{transmitted}=\alpha_i\dot{\delta_i} -P_{max}sin(\Delta\delta_{ij})=P_{transmitted}-P_{dissipated}$. It can be seen that the collective acceleration of generators is dependent on the difference of mechanical and electrical power inputs. As shown in Figure \ref{fig:equal_area_martens}, difference in electrical-mechanical inputs either accelerate or decelerate the generators to achieve equilibrium. The generators accelerate when mechanical power is higher than the transmitted electrical power (i.e., $P_{m,i}>P_{transmitted}$) and decelerate when electrical power is higher (i.e., $P_{m,i}<P_{transmitted}$). This is due to the fact that, the difference in the power gives rise to the rate of change of accumulated power ($P_{accumulated}$) in the rotor masses. The change in accumulated power and generator inertia results in effective change in rotor angles, thereby acceleration/deceleration and vice-versa. The coupled set of generators happen to achieve steady-state, when mechanical power of the generator from turbines match the transmitted electrical power. This can be visualized as creation of a `sink' node, where the system tries to drive itself in order to achieve stability. Thus, any generator starting from a rotor angle $ \in [\delta_0,\delta_c,\delta_m]$, will try to move towards $\delta_0$ (or stability). On the other hand, $\delta_m$ happens to be critically stable and a small perturbation towards $\pi$ radians can render increase in rotor angles due to effective acceleration of generators. In such case, a `source' node is formed at $\pi$ radians, as shown in Figure \ref{fig:equal_area_martens}. 
\par Generators are mechanical devices that exhibit high inertia and hence require time to achieve stability, once perturbed from its equilibrium. The natural speed (angular speed) of rotation $\omega_i$ and generator inertia $J$ along-with accumulated power $P_{accumulated}$ in turn defines the rate at which stability is achieved or if system becomes unstable. As shown in Figure \ref{fig:equal_area_martens}, although $\omega_i$ and $J$ are constants if generator achieves stability via route 2 has higher $P_{accumulated}$ and is highly susceptible to instability as compared to route 1. We show this using circle plot and thereby its relevance to `$\pi$', `transmitted' and `incoherence' state as mentioned in previous sections. Thus, it can be inferred that a `source' node imparts instability whereas `sink' node stabilizes a power setup. These inferences can be easily made using CC-Kuramoto model as these nodes are pretty evident. In the next section, we do a bifurcation analysis on CC-Kuramoto model in order to understand the effect of system parameters on system stability.

\subsection{Case Study: Bifurcation Analysis}

\begin{figure}[t!]
\centering
\begin{tabular}{c}
\includegraphics[height=5.0cm,width=8.6cm]{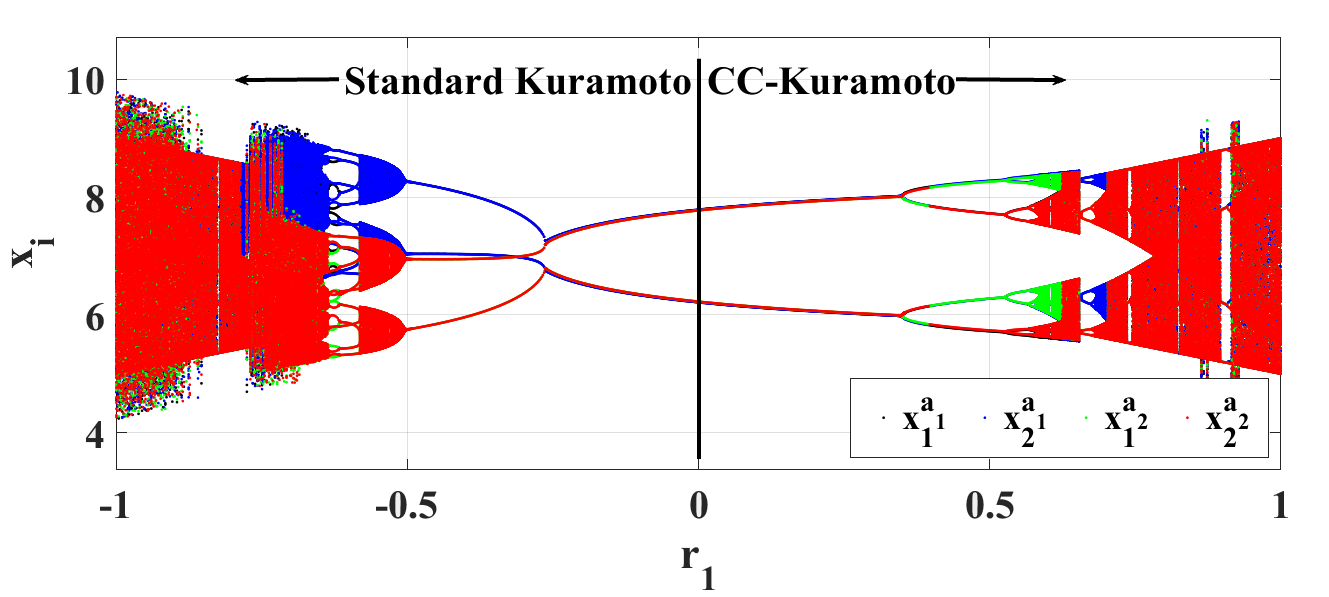}\\
(a)\\
\includegraphics[height=5.0cm,width=8.6cm]{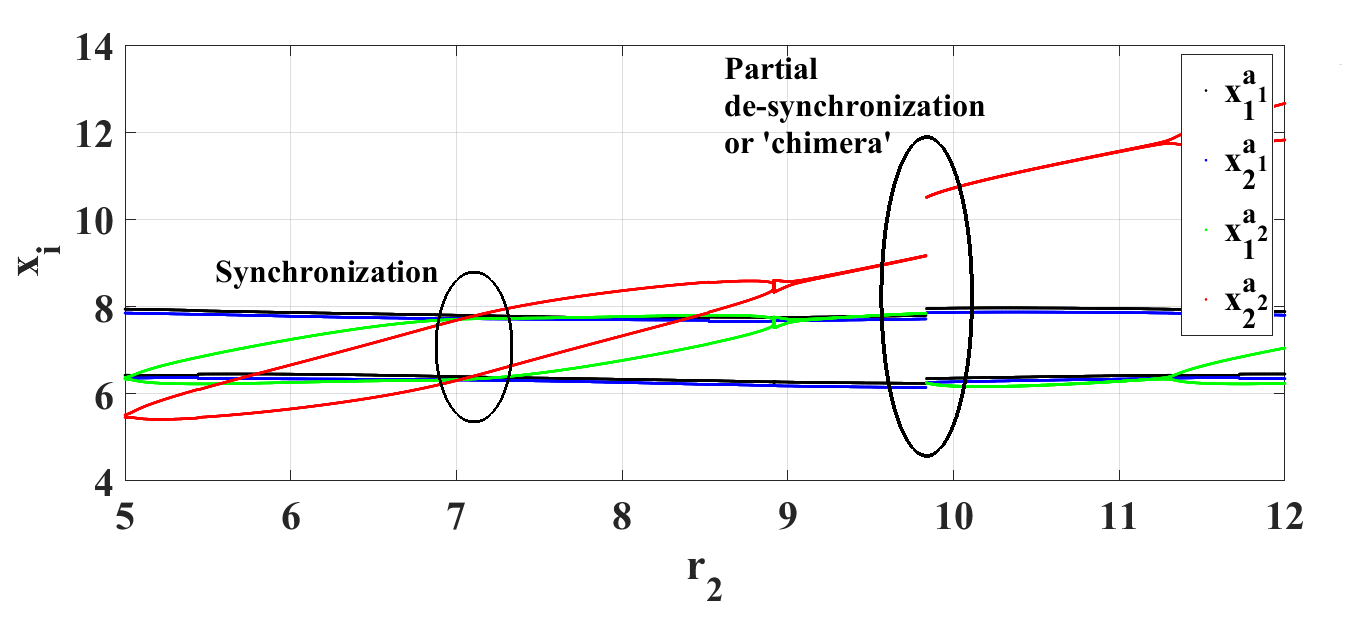}\\
(b)
\end{tabular}
\caption{Bifurcation Analysis of CC-Kuramoto model. (a) Analysis of system stability with variation in interarea coupling constants. (b) Stability analysis of the model by varying the natural frequencies in certain areas.}
\label{fig:bifurcation_analysis}
\end{figure}

\subsubsection{Using Equilibrium Points}

\par Next, we rewrite \eqref{interarea} in order to study the bifurcation in power system using CC-Kuramoto model. Let $\delta_i^{a_c}=x_i^{a_c}, \dot{\delta}_i^{a_c}=y_i$, and $r_1$ be the parameter of bifurcation on interarea coupling.

\begin{equation}
\begin{split}
0=\omega_i-\alpha_i y_i \ &+\sum_{j\neq i,j=1}^{p}k_{ij}sin(x_j^{a_1} - x^{a_1}_i)\\ &-\sum_{j=p+1}^{n}r_1 \ sin(x^{a_2}_j - x^{a_1}_i),\\
0=\omega_i-\alpha_i y_i \ &+\sum_{j\neq i,j=p+1}^{n}k_{ij}sin(x_j^{a_2} - x^{a_2}_i)\\ &-\sum_{j=1}^{p}r_1 \ sin(x^{a_1}_j - x^{a_2}_i).
\end{split}
\label{bifur_eq}
\end{equation}

\begin{figure*}[t!]
\centering
\begin{tabular}{c}
\includegraphics[height=7.5cm,width=16cm]{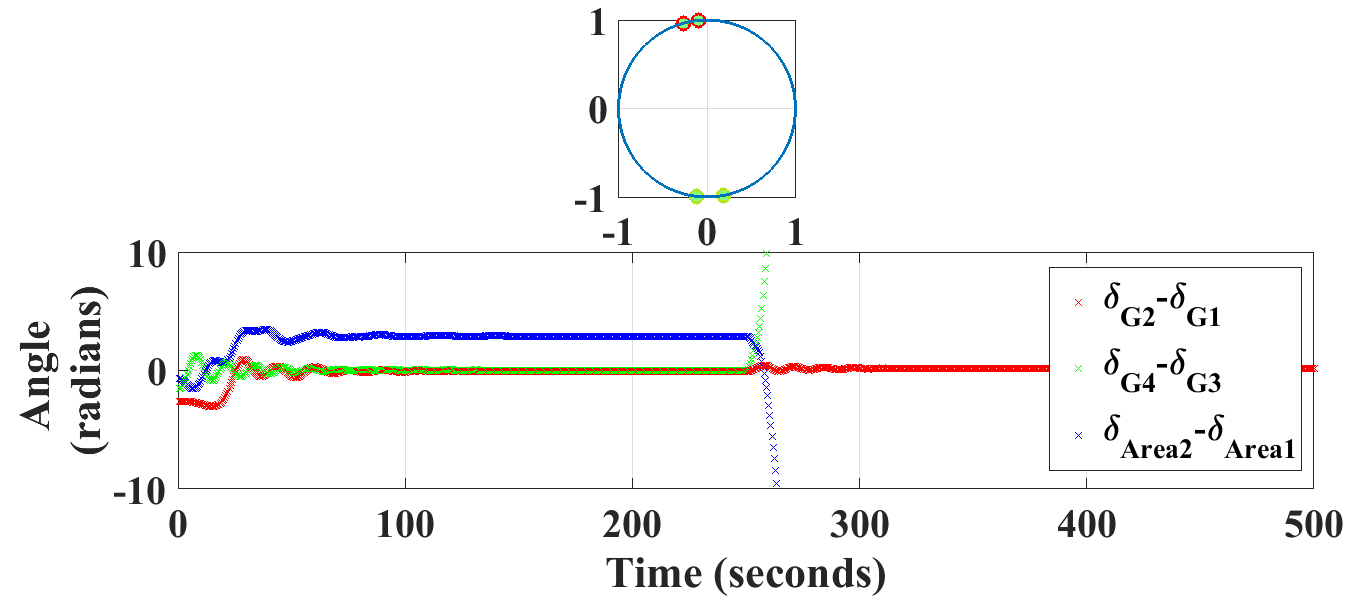}\\
(a)
\end{tabular}
\begin{tabular}{cc}
\includegraphics[height=4.2cm,width=8cm]{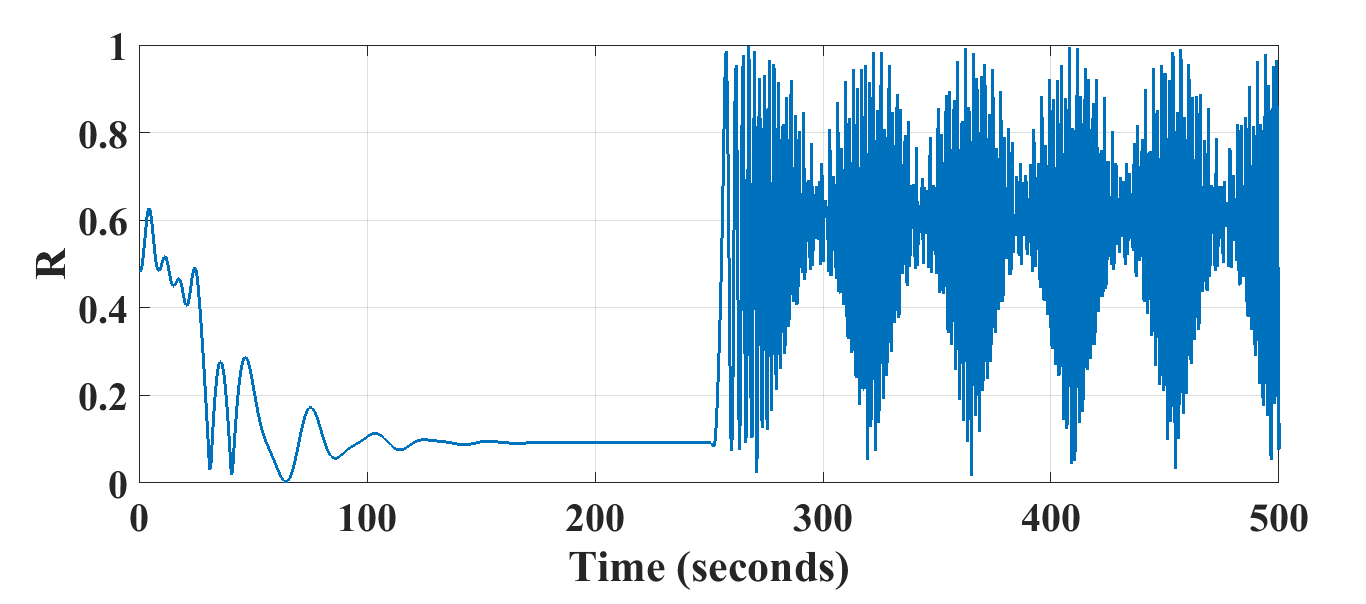}&
\includegraphics[height=4.2cm,width=8cm]{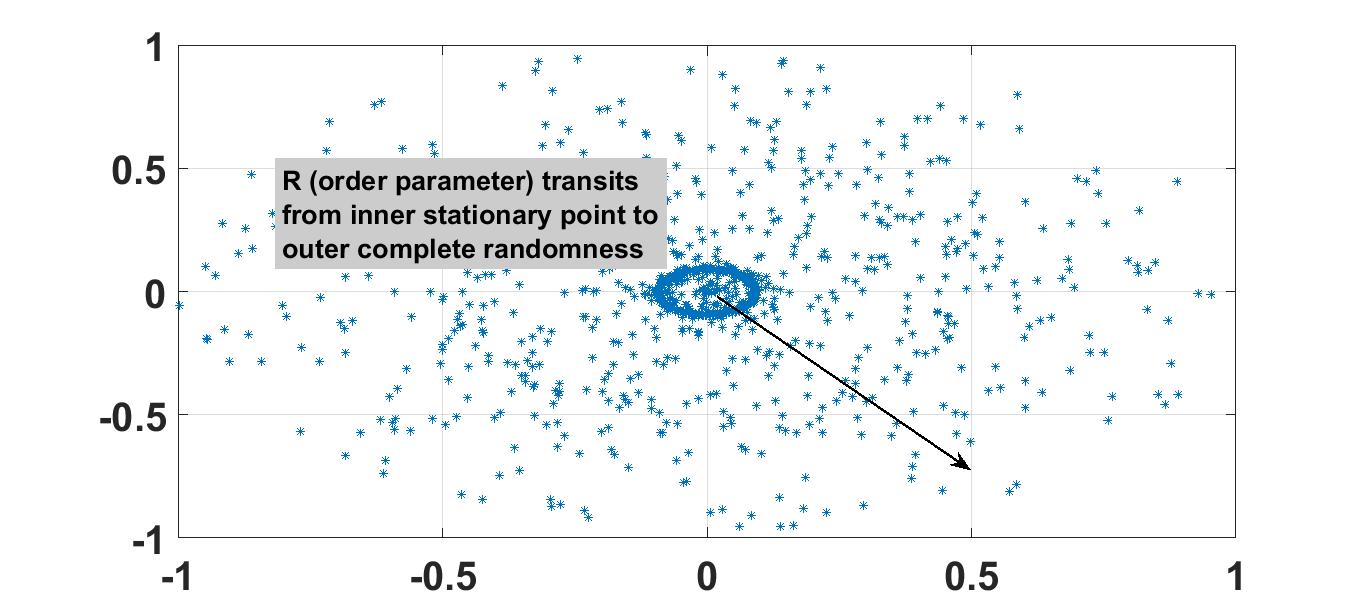}\\
(b)&(c)
\end{tabular}
\caption{Simulation of `chimera' behavior in a system of coupled generators. (a) Circle plot against Time series plot showing loss of synchronicity at time $\sim 250$ seconds. (b) Order parameter ($R=\frac{1}{n}\sum^{n}_{j=1}e^{i\delta_j}$) plot against time. (c) Polar plot of order parameter.}
\label{fig:chimera}
\end{figure*}

\par With the same setup of two-area four-machine system, parameter $r_1$ (i.e., interarea coupling) is varied in the range $r_1 \in [-1,1]$ with homogeneous natural frequencies (i.e., $\omega_i=\omega$). Under this setting, we solve for equilibrium points of $\dot{x}_i^{a_c}=\dot{y}_i=f(x_i^{a_c},y_i)=0$ or $\ddot{x}_i^{a_c}=0$ and thereby fixed points of the dynamical equations \eqref{bifur_eq}. The results were obtained using MATLAB and are shown in Figure \ref{fig:bifurcation_analysis} (a). It can be seen that interarea coupling of $0.5 \geq r_1 \geq -0.5$ achieves period doubling, although the solutions are constant. Beyond $-0.5>r_1>0.5$ CC-Kuramoto model achieves chaotic behavior. This can be visualised as complete loss of synchronism and thereby chaos. 

\begin{equation}
\begin{split}
0=\hat{\omega}_i-\alpha_i y_i \ &+\sum_{j\neq i,j=1}^{p}k_{ij}sin(x_j^{a_1} - x^{a_1}_i) \\ &-\sum_{j=p+1}^{n}k_{ij} \ sin(x^{a_2}_j - x^{a_1}_i),\\
0=\hat{\omega}_i-\alpha_i y_i \ &+\sum_{j\neq i,j=p+1}^{n}k_{ij}sin(x_j^{a_2} - x^{a_2}_i)
\\ &-\sum_{j=1}^{p}k_{ij} \ sin(x^{a_1}_j - x^{a_2}_i),
\end{split}
\label{bifur_eq2}
\end{equation}

\par For the next case, $\omega^{a_2}_4$ is varied in a range $\omega^{a_2}_4=r_2 \in [5,12]$rad/s keeping coupling parameter constant, showing synchronization at $r_2=7$rad/s and leaves synchronicity at $r_2=10$rad/s showing chimera behavior. We solve for \eqref{bifur_eq2}, where $\hat{\omega}_i=[\omega^{a_1}_1,\omega^{a_1}_2,\omega^{a_2}_3,r_2]^T$. This scenario can be interpreted as a gradual overload of one of the generators from two areas leading to de-synchronization in one area, whereas other area remains synchronized (refer Figure \ref{fig:bifurcation_analysis} (b)). Further, using circle and time-series plots for angular separations (as shown in Figure \ref{fig:chimera} (a)), the same partial de-synchronization is observed. These can be inferred as islanding of power network through circuit breakers to avoid the impact of excessive overloading of generators in a neighbouring area (and hence blackouts or cascaded failures \cite{dey2016impact}). To summarize, the heterogeneity was introduced by increasing frequencies $\omega_i$ incrementally in one area, while keeping parameters of other area constant \big (i.e., $\omega^{a1}_1,\omega^{a1}_2,\omega^{a2}_3 \in g_1(\omega);\ \omega^{a2}_4 \in r_2 g_1(\omega)=g_2(\omega), r_2 \in \mathbb{R}$; $g_1,g_2$ being frequency distributions\big). As seen from Figure \ref{fig:chimera} area experiencing incremental perturbations loose synchronicity whereas other area remains unaffected, emulating blackout conditions with islanding. In order to measure loss of synchronicity, we use order parameter $R=\frac{1}{N}\sum^{N}_{j=1}e^{i\theta_j}$ as shown in Figure \ref{fig:chimera} (b), (c).

\subsubsection{Using Eigen value analysis}

\par In this subsection we provide an eigen value based justification for bifurcation phenomena observed in previous subsection. For instance, consider \eqref{axonal3} and let $\lambda_{ap}, \lambda_{ip}$, $\lambda_{inc}$ be eigenvalues of anti-phase, in-phase and incoherent dynamics respectively. Where

\begin{equation}
    \begin{aligned}
       \lambda = \begin{cases}
\lambda_{ap}<0 & \text{if} \ \Phi_{ij}=m(\pi)\\
\lambda_{ip}>0 & \text{if} \ \Phi_{ij}=0, m(2\pi)\\
\lambda_{inc}=0 & \text{if} \ \Phi_{ij}=m(\pi/2), 
\end{cases}
    \end{aligned}
    \label{eigenanalysis}
\end{equation}

\par $m \in \mathbb{Z}$. Now, since $\lambda_{ap}=-\lambda_{ip}$, these two nodes exchange stability through $\lambda_{inc}$ and hence following can be concluded. (i) anti-phase and in-phase modes have converse stabilities and are never stable simultaneously, (ii) these critical modes swap stabilities at incoherence state and (iii) if either of anti-phase or in-phase states are stable incoherence state must be unstable and vice-versa. Particularly, in power systems these states rest in anti-phase (unstable), in-phase (stable) and chimera (partially stable) states.

\section{Conclusions}
\par In this study, a mathematical model for interarea oscillations is proposed using Kuramoto-type framework with its analogy in power grids. It is shown how these oscillations can be visualized in a `conformist-contrarian' form to better understand interarea oscillations. Validity of the choices has been justified using analogy of spring coupled pendulums. In order to verify the model, a standard four generator power system was considered from the literature. Simulations were performed in MATLAB and results were verified and validated. The proposed model is used to investigate various phenomena like spatial/temporal chimera \cite{abrams2004chimera} and spontaneous failures in power systems \cite{nerc9209}. 

\section*{Acknowledgements}
The authors would like to thank Prof. S. D. Varwandkar for his valued discussions and feedbacks towards the outcome of this work.

\end{document}